\documentstyle[aps,preprint,psfig]{revtex}
\begin{document}
\title{\bf Condensation of Hard Spheres Under Gravity}

\author{Daniel C. Hong\thanks{E-mail address: 
   dh09@lehigh.edu} }

\address{ Department of Physics, Lewis Laboratory, Lehigh University, 
Bethlehem, Pennsylvania 18015}

\date{\today}
\maketitle

\begin{abstract}
Starting from Enskog equation of hard spheres of mass m and diameter D
under the gravity g,
we first derive the exact equation of motion
for the equilibrium density profile at a temperature T 
and examine its solutions via the gradient expansion.  The solutions
exist only when $\beta\mu \le \mu_o \approx 21.756$ in 2 dimensions
and $\mu_o\approx 15.299$ in 3 dimensions, where $\mu$ is the dimensionless
initial 
layer thickness and $\beta=mgD/T$.
When this inequality breaks down, 
a fraction of particles condense from the bottom up to
the Fermi surface.
\vskip 1.0 true cm
\end{abstract}

\noindent PACS numbers: 05.20-y, 81.35+k, 05.20.Dd, 05.70.Fh
\vskip 1.0 true cm
Granular materials are basically a collection of hard spheres that interact
with each other via hard sphere potential [1].  For this reason, many of the
properties of excited granular materials may be understood from the
atomistic view of molecular gases, in particular
from the view point of kinetic theory [2].   There are, however,
several distinctions between
molecular gases and granular materials: First, granular materials are
macroscopic particles with finite diameter, and thus they cannot be
compressed indefinitely.  Second, the gravity plays an important role in
the collective response of granular materials to external stimuli, largely
because of the ordering of grains induced by the gravity.
For example, one of the notable characteristics of the
excited granular materials in a confined system under gravity  
is the appearance of a thin
boundary layer near the surface that separates a fluidized region from a 
rigid solid region.  This has been known for long time as is 
evidently seen in shearing experiments [3], avalanches [4], 
and grains subjected to
weak excitations [5,6].  In this limit, those grains 
in a solid region are effectively
frozen, and thus do not participate in dynamical processes.  Now,  
the kinetic theory relies on two particle
collision dynamics, and has been applied to systems where all the
granular particles are in motion colliding with each other.  
But if a portion of grains are frozen and remain largely motionless, 
the kinetic theory or in general the continuum theory may
pose some problems.
Consider for example a system of strongly excited
granular particles under gravity,
where all the grains undergo collisions and thus the
kinetic theory is valid.  If we decrease the strength of excitation, then
the particles at the bottom will freeze themselves, and the boundary
layer will develop at the top. The question we address in this Letter is:
How does the kinetic theory describe such process?  
\vskip 0.2 true cm
In a recent paper [6], it has been demonstrated that
the granular statistics in the presence of 
gravity does not follow the usual Boltzmann statistics as in molecular
gases, where all the particles are dynamically active,
but a new Fermi statistics, where most of the particles
are effectively frozen and only a portion of particles
near the surface participate in the dynamical process.  
This is due to the excluded volume
effect and the ordering of potential energy by gravity, and the mechanism
associated with this Fermi statistics is similar to that of 
the Fermi gas in a metal.
The existence of a thin boundary layer in granular materials should be 
viewed from such perspective.
\vskip 0.2 true cm
Our specific objective of this Letter 
is to use the kinetic theory, in particular the Enskog equation 
of hard spheres of mass m and diameter D, to explore
whether or not the kinetic theory can describe the 
cross over from Boltzmann to Fermi
statistics and if so, under what conditions it occurs.  
Our particularly interesting discovery is that
the prediction of the Enskog equation is only valid when 
$\beta\mu \le \mu_o $, where $\mu$ is the dimensionless initial layer
thickness of the granules(or the Fermi energy),
$\beta=mgD/T$  with T the temperature, and
the critical value, $\mu_o$, is determined to be
$\mu_o=21.756$ in 2d and $\mu_o=15.299$ 
in 3d.  When this inequality is violated,
Enskog equation does not conserve the particles, and the missing particles
condense from the bottom up to the Fermi surface.
This way, the hard sphere Enskog gas appears to contain
the essence of Fermi statistics and Bose condensation.
\vskip 0.3 true cm
The starting point of our investigation is the
Enskog's kinetic equation for elastic hard spheres [7] of mass m and diameter D
in the presence of 
gravity:

$$\frac{\partial f}{\partial t} + {\bf v}\bullet\nabla f - mg \frac{\partial
f}{\partial v_z} = J_E \eqno (1)$$

where the Enskog's collision operator is given by:

$$J_E = D^2\int d^3v_1\int_+d^2e{\bf e}\bullet{\bf g}[
f({\bf r,v'})f({\bf r}+D{\bf e,v_1'})\chi({\bf r}+\frac{1}{2}D{\bf e})
-f({\bf r,v})f({\bf r}-D{\bf e, v_1})\chi({\bf r}-\frac{1}{2}D{\bf e})
\eqno (2)$$

Here, $({\bf v},{\bf v}_1)$  and $({\bf v'},{\bf v}_1')$ are the 
velocities of two colliding particles before and after the collision, and
${\bf r}$ and ${\bf r}'$ are the positions of the two particles when they are
in contact, 
{\bf e} is the unit vector in the direction of {\bf r}-{\bf r'},
{\bf V} is the relative particle flux given by 
${\bf V}={\bf v}-{\bf v_1}$
and ${\bf V'} = {\bf v}'-{\bf v_1}'$ [2].  The + sign means that the
integration should be carried out with the restriction that 
$ 0\le {\bf e}\bullet{\bf
V}$. From the geometry it is easy to show
${\bf V}' = {\bf V} - 2{\bf e}{\bf e}\bullet{\bf V}$ and $|{\bf V}|
'=|{\bf V}|$.  The correlation function, $\chi$, is defined as
$ F_2({\bf r},{\bf v}; {\bf r}_1,{\bf v}_1) = F_1({\bf r},{\bf v})
F_1({\bf r},{\bf v}_1)\chi$,
where $F_2$ and $F_1$ are the two and one particle distribution function
respectively.
Molecular chaos assumption in the usual Boltzmann collision
operator sets $\chi=1$.  In the case of dense gases,
Enskog [2,7] assumed that $\chi$ might be given
by the equilibrium two point 
correlation function estimated
at the contact point.   
At equilibrium, we expect that the factorization of
the space and velocity in the distribution function is valid. Thus, we set:  
$ f({\bf r},{\bf v};t=\infty) = G({\bf r}) \phi({\bf v})$.  This is equivalent
to the separation of the configurational statistics from the kinetics.
In equilibrium statistics of elastic granular materials, 
$\phi({\bf v})$ is expected to be Gaussian, and the
energy conservation requires that $\phi({\bf v}_1')\phi({\bf v}')=
\phi({\bf v}_1)\phi({\bf v})$.
We note that $\phi({\bf v})$ is normalized
to one: $\int d^3{\bf v}\phi({\bf v}) =1$.  
In the steady state, 
the left hand said of eq.(1) then becomes:

$$v_z\phi({\bf v})[\frac{\partial G}{\partial z} + \frac{mg}{T} G(z)]
\eqno (3)$$

In order the compute the right hand side, we first notice that
the collision operator $J_E$(eq.(2)) is given by:

$$ J_E = D\phi({\bf v})G({\bf r})[\int d^2{\bf v}\phi({\bf v}_1)]\int_{+}
d\theta{\bf e}[\chi({\bf r}-\frac{1}{2}D{\bf e})G({\bf r}-D{\bf e}) -
\chi({\bf r}+{\frac{1}{2}}D{\bf e}) G({\bf r} + D{\bf e})]{\bf e}
\bullet{\bf V} \eqno (4)$$

We now first compute the right hand side of eq.(1)
under the factorization assumption.
In this case, a care must be taken because 
the integral should be performed with the constraint 
${\bf e}\bullet{\bf V} \ge 0$.  However, 
if we change ${\bf e}$ to $-{\bf e}$ of the second term in the Enskog operator
$J_E$(eq.(4)), we find it becomes:

$$
-\int_{{\bf e}\bullet{\bf V}\ge 0}
d\theta \chi({\bf r}+{\frac{1}{2}}D{\bf e})
G({\bf r} + D{\bf e}){\bf e}\bullet{\bf V}
=\int_{ {\bf e}\bullet{\bf V}\le 0} d\theta
{\bf e}\bullet{\bf V}\chi({\bf r}-{\frac{1}{2}}D
{\bf e})G({\bf r} - D{\bf e})\eqno (5)$$

Hence, we can now {\it remove} the restriction ${\bf e}\bullet
{\bf V}\ge 0$ in eq.(4)
and integrate over the {\it whole} space.  After some algebra, we obtain:
$$
J_E = D\phi({\bf v})G({\bf r})[\int d^2{\bf v}_1\phi({\bf v}_1)]
{\bf v}\bullet{\bf I}\eqno (6)$$

where 
$${\bf I} 
= \int_{all space}
 d\theta{\bf e}\chi({\bf r}-{\frac{1}{2}}D{\bf e})
G({\bf r}-D{\bf e}) \nonumber$$
$$= {\frac{1}{2}}\int_{all space} d\theta
{\bf e}{\bf e}[\chi({\bf r}-{\frac{1}{2}}D
{\bf e})G({\bf r}-D{\bf e})-\chi({\bf r}+{\frac{1}{2}}D{\bf e})G({\bf r}+
D{\bf e})] \eqno (7)$$

In obtaining (7), we utilized the fact that $\int d^2{\bf v}_1\phi({\bf v}_1)
{\bf v}_1 = 0$ by symmetry.
Our next step is to compute {\bf I}.  To this end, we first rewrite {\bf I}:

$${\bf I} 
= \int d\theta{\bf e}\chi({\bf r}-{\frac{1}{2}}D{\bf e})
G({\bf r}-{\frac{1}{2}}D{\bf e}) = -\int d\theta{\bf e}\chi({\bf r} + 
{\frac{1}{2}}D{\bf e})G({\bf r}+{\frac{1}{2}}D{\bf e}) \nonumber $$
$$= {\frac{1}{2}}\int d\theta{\bf e}[\chi({\bf r}-{\frac{1}{2}}D{\bf e})
G({\bf r}-D{\bf e}) - \chi({\bf r}+{\frac{1}{2}}D{\bf e})G({\bf r}+
{\frac{1}{2}}D{\bf e})]\eqno (8)$$

Notice that $I_x$ (and $I_y$) will
vanish by symmetry and only the vertical component, $I_z$, survives. 
Using ${\bf e}=(\sin\theta, \cos\theta)$ in 2d, we find:

$$I_z = {\frac{1}{2}}
\int_{0}^{2\pi}d\theta cos\theta[\chi(z-{\frac{1}{2}}Dcos\theta)G(z-Dcos\theta)
-\chi(z+{\frac{1}{2}}Dcos\theta)G(z+Dcos\theta)] \eqno (9)
$$

At equilibrium, $\phi({\bf v};T) = \phi(m{\bf v}^2/2T)$.
We now put this functional form along with $G(z;T)=G(mgz/T)$
to the right hand side of (1) 
and cancel
$v_z\phi({\bf v})$ term.  Next,
in a free volume theory, particles are confined in a cage.  Hence, if we use
a simple cubic lattice as a basic lattice, the closed packed volume fraction
$\rho_c = N/V = N/D^2N = 1/D^2$.  
If we define the dimensionless density
$\phi(z) =G(z)/\rho_c=D^2 G(z)$ or 
$\phi(\zeta,\beta)=D^2G(z)$ 
with $\zeta=z/D$,
we then obtain the exact dimensionless
equation of motion for $\phi(\zeta,\beta)$:

$${\frac{d\phi(\zeta)}{d\zeta}} + \beta\phi(\zeta) = \phi(\zeta) I_{\zeta}
(\zeta)
\label{(10)}\eqno (10)$$
where

$$I_{\zeta}(\zeta) = {\frac{1}{2}}
\int_{0}^{2\pi}d\theta[\chi(\zeta-{\frac{1}{2}}cos\theta)
\phi(\zeta-cos\theta) - \chi(\zeta+{\frac{1}{2}}cos\theta)\phi(\zeta
+cos\theta)]
\eqno (11)$$

For 3d,the corresponding equation
for the density $\phi(\zeta)=D^3G(z)$ is given by:
$$ \frac{d\phi}{d\zeta} + \beta\phi = \phi I_{\zeta}(\zeta) \eqno (12)$$
with
$$ I_{\zeta}(\zeta)=
\pi\int^{\pi}_0 d\theta sin\theta cos\theta [\chi(\zeta-cos\theta/2)
\phi(\zeta-cos\theta) - \chi(\zeta+cos\theta/2)\phi(\zeta
+cos\theta)] \eqno (13)$$
Several forms for the equilibrium correlation function $\chi$ have been 
proposed, but we use the following widely used forms:
For 2d, we use the form proposed by Ree and Hoover [8]:
$\chi(\phi) =(1-\alpha_1\phi + \alpha_2\phi^2)/(
(1-\alpha\phi)^2$, while for
3d, we use the form suggested by Carnahan and Starling [9]:
$\chi(\phi) = (1-\pi\phi/12)/(1-\pi\phi/6)^3$

Since the total number of particles, $N$, remain fixed,
the following normalization condition should be satisfied for both 2d and 3d.

$$\int_o^{\infty}d\zeta \phi(\zeta;\beta) = \mu \eqno (14)$$

\noindent where $\mu \equiv N/\Omega_x$ (or $\mu\equiv 
N/\Omega_x\Omega_y$ in 3d) is 
the Fermi energy [6] and $\Omega_x,\Omega_y$
are the degeneracies along the x and y axis.
We now perform the gradient expansion of (11) and (13) and retain only the
terms to first order in $d\chi/d\zeta$.  We find:

$$\frac{d\phi}{d\zeta} + \beta\phi = -\frac{\pi}{2}\phi
[\frac{d\chi}{d\zeta}\phi + 2\chi\frac{d\phi}{d\zeta}] 
\qquad (2d) \eqno (15a)$$

$$\frac{d\phi}{d\zeta} + \beta\phi = -\frac{2\pi}{3}\phi[2\frac{d\phi}
{d\zeta}\chi + 
\phi\frac{d\chi}{d\zeta}]\qquad (3d)\eqno (15b)$$

The solutions are readily obtained.  For 2d, we find:
$$-\beta(\zeta-\bar\mu) = ln\phi + c_1\phi + c_2log(1-\alpha\phi)
+ c_3/(1-\alpha\phi) + c_4/(1-\alpha\phi)^2\label{(15)}\eqno (16a)$$

$$ \beta\bar\mu = ln\phi_o + c_1\phi_o + c_2ln(1-\alpha\phi_o)
+c_3/(1-\alpha_o) + c_4/(1-\alpha\phi_o)^2 \eqno (16b)$$

$$\beta\mu = \phi_o + c_1\phi_o^2/2 + c_2(\phi_o + ln(1-\alpha\phi_o)/\alpha)
+c_3ln(1-\alpha\phi_o)/\alpha $$
$$ -(c_4/\alpha)[1/(1-\alpha\phi_o)-1] + c_3\phi_o/(1-\alpha\phi_o)
+ c_4\phi_o/(1-\alpha\phi_o)^2 \eqno (16c)$$
where
$\phi_o$ is the density at $\zeta=0$, and
$c_1 = 2\alpha_2/\alpha^2\frac{\pi}{2} \approx 0.0855$,
$c_2 = -\frac{\pi}{2}
(\alpha_1 - 2\alpha_2/\alpha)/\alpha^2\approx 0.710$,
$c_3=-c_2$,$c_4= \frac{\pi}{2}
(1-\alpha_1/\alpha + \alpha_2/\alpha^2/)\alpha \approx 1.278$.
For 3d, we find:

$$-\beta(\zeta-\bar\mu) = ln\phi -1/(1-\alpha\phi)^2 + 2/(1-\alpha\phi)^3
\eqno (17a)$$

$$\beta\bar\mu = ln(\phi_o) -1/(1-\alpha\phi_o)^2 + 2/(1-\alpha\phi_o)^3
\eqno (17b)$$

$$ \beta\mu = \phi_o - \frac{2\phi_o}{1-\alpha\phi_o} + 
\frac{2\phi_o}{(1-\alpha\phi_o)^3}\eqno (17c)$$

\noindent where $\alpha=\pi/6$.
For given values of $\beta$ and $\mu$, 
$\phi_o \equiv \phi(\zeta=0)$ will be determined
by eq.(16c) and (17c).  However, 
since the right hand sides are
monotonically increasing functions for $\phi_o $,
$\beta\mu$ must have
the upper bound $\mu_o$; namely, $\mu_o=21.756$ in 2d and $\mu_o=15.299$ in 3d,
which are the values obtained by setting $\phi_o=1$
in the right hand side of (16c) and (17c).  Considering the
fact that both the temperature $T$ and
the Fermi energy $\mu$ are {\it arbitrary} control parameters, the existence
of such bounds is a puzzle: if $\beta\mu$ is less than $\mu_o$, then
the density profiles given by Eq.(16a) and (17a) are well determined, but
if $\beta\mu$ is greater than $\mu_o$, then $\phi_o$ must be one,
and the particle conservation breaks down, namely

$$\int_o^1d\phi \zeta(\phi)=\int_0^{\infty}d\zeta\phi(\zeta) 
\equiv \mu_o/\beta < \mu \eqno (18)$$

The central
question is: where does the rest of the particles go?  In order to gain
some insight into this question, consider first the case of point particles
under gravity, in which case the density profile is given by: $\rho(\zeta)=
\rho(0)
exp(-mg\zeta/T)$.  If we put more particles into the system, we simply
need to increase $\rho(0)$ because the point particles can be compressed 
indefinitely, and the profile
simply shifts to the right.  We now replace these point particles with
hard spheres, which cannot be compressed indefinitely.
Suppose we start from a high temperature
where all the particles are active.  We then slowly decrease the temperature 
to suppress the thermal motion.  At a certain temperature, the freezing of the
particles will occur from the bottom [11], which will then spread out as
the temperature is lowered down further, 
until at T=0 all the particles are frozen.
Note that the frozen particles in the 
closed packed region behave like a solid.  
Such observation helps us to resolve the puzzle associated with the
disappearance of particles, which must then condense from the bottom up
to the lower part of the fluidized layer.
We term this
surface, which separates the frozen or a closed packed region with
$\phi=1$ from the fluidized region with $\phi <1$,
the Fermi surface.  The location of the Fermi surface, $\zeta_F$,
is determined by
the amount of the missing particles, namely, $\zeta_F=\mu-\mu_o/\beta$.
For nonzero $\zeta_F$, we must put the
missing particles below the Fermi surface and
shift the bottom layer from $\zeta=0$ to $\zeta_F$.  Such modified 
profile for $\beta = 1$ and $\mu=100$ in 2d is shown in Fig.1.  In comparison,
in Fig.1 is also shown the density profile for $\beta=10$ and
$\mu=100$ for which $\beta\mu \le \mu_o$.  We conclude
that while the density profile obtained this way
is not exactly the same as the Fermi profile,
the essence of the Fermi statistics, namely the effect of
excluded volume interactions, sets in when $\beta\mu\ge\mu_o$.  
The condensation of particles from the bottom in one
dimensional vibrating bed [12] and
the clustering of particles near the bottom wall in
two dimensional experiments [13] appear to be a strong confirmation of this
scenario, which seems to be reminiscent of the Bose condensation of particles
into the ground state.

We conclude with a few remarks.  First, it remains to
be seen whether the feature observed in this paper for hard spheres
persists in the presence of dissipation, namely
when particles collide inelastically.  
In this case, there
are some evidences that the velocity distribution function is 
not Gaussian for strong dissipation,
 and the factorization assumption may not be valid.
The future studies must
focus on the accurate determination of this velocity profile,
based on which an extension of the present analysis must be carried out.
Second, we may define the freezing temperature 
$T_c$ as the point where the particle conservation breaks down, namely
$$T_c= mgD\mu/\mu_o \eqno (19)$$
which may be tested experimentally presumably by 
using the relation between T and
the vibration strength $\Gamma$ [6], or by Molecular Dynamics simulations.
Finally, we point out that eqs. (15a) and (15b)
can also be obtained
from the force balance eq. for the pressure P:$dP/dz = -\rho g $
with P obtained by the virial expansion of hard sphere gas in the absence of
gravity [10]: $P/T\rho = 1 + A_d\rho D^d\chi(D)/2d$,
where $A_d=4\pi$ for 3d and $A_d=2\pi$ for 2d.

\vskip 1.0 true cm
The author is particularly grateful to A. J. McLennan
for numerous discussions on kinetic theory and for
several incisive suggestions over the course of this work.  The author also
wishes to thank H. Hayakawa for helpful discussions, 
S. Luding for ref. [12] and
Joe Both and Paul Quinn for checking some of the algebras.

\vskip 1.0 true cm
\vfill\eject
\noindent {\bf References}
\vskip 0.3 true cm
\noindent [1] H. Jaeger,
S.R.Nagel and R.P.Behringer, Physics Today, {\bf 49}, 32 (1996),
Rev.Mod.Phys. {\bf 68}, 1259 (1996);
H.Hayakawa, H. Nishimori, S. Sasa and Y-h. Taguchi, Jpn. J. Appl. Phys. Part 1,
{\bf 34}, 397 (1995) and references therein.
\vskip 0.2 true cm
\noindent [2] J. Jenkins and S. Savage, J. Fluid. Mech. {\bf 130}, 197 (1983);
S. Chapman and T. G. Cowling, {\it The Mathematical
Theory of Nonuniform Gases} (Cambridge, London, 1970); J.A. McLennan,
{\it Introduction to Non-Equilibrium Statistical Mechanics}, 
Prentice Hall (1989).
\vskip 0.2 true cm
\noindent [3] D. M. Hanes and D. Inman, J. Fluid. Mech. {\bf 150}, 357 (1985);
S. Savage and D. Jeffrey, {\it ibid}, {\bf 110}, 255 (1981)
\vskip 0.2 true cm
\noindent [4] For example, see: J.P. Bouchaoud, M.E. Cates, R. Prakash, and
S.F.Edwards, J. Phys. France {\bf 4}, 1383 (1994)
\vskip 0.2 true cm
\noindent [5] E. Clement and J. Rajchenbach, Europhys. Lett. {\bf 16}, 133
(1991); J.A. Gallas, H. Herrmann, and S. Sokolowski, Physica A, {\bf 189}, 
437 (1992)
\vskip 0.2 true cm
\noindent [6]
 H. Hayakawa and D. C. Hong, Phys. Rev. Lett., {\bf 78}, 2764 (1997)
\vskip 0.2 true cm
\noindent [7] D. Enskog, K. Sven. Vetenskapsaked, Handl. {\bf 63}, 4 (1922);
\vskip 0.2 true cm
\noindent [8] F. R. Ree and W. G. Hoover, J. Chem. Phys. {\bf 40}, 939 (1964)
\vskip 0.2 true cm
\noindent [9] N.F. Carnahan and K.E. Starling, J. Chem. Phys. {\bf 51}, 635 
(1969).
\vskip 0.2 true cm
\noindent [10] J.-P. Hansen and McDonald, Theory of Simple Liquids,
2nd Edition (Academic, London, 1986).
\vskip 0.2 true cm
\noindent [11] The freezing here does not mean the complete suppression of
thermal motion.  It means the suppression of
the translational motion for those particles in a closed packed regime.
\vskip 0.2 true cm
\noindent [12] S. Luding, 'Models and simulations of granular materials,'
Ph.D thesis, Albert-Ludwigs University, Germany (1994).  See Fig.19.
\vskip 0.2 true cm
\noindent [13] A. Kudrolli, M. Wolpert and J.P. Gollub, Phys. Rev. Lett. 
{\bf 78}, 1383 (1997)
\newpage
\noindent {\bf Figure Caption}
\vskip 0.3 true cm
Fig.1.  The crossover from Boltzmann to Fermi statistics
as the temperature is lowered.  For a given Fermi energy, $\mu=100$, 
the density profiles are shown as a function of dimensionless height $\zeta$
for $\beta = 1/10$ and $\beta=1$(dotted line) for
the two dimensional Enskog gas.  For
$\beta=1$, grains freeze from the bottom up to the Fermi surface, and
only those grains near the Fermi surface participate in the dynamical process.
\newpage
\centerline{\hbox{
\psfig{figure=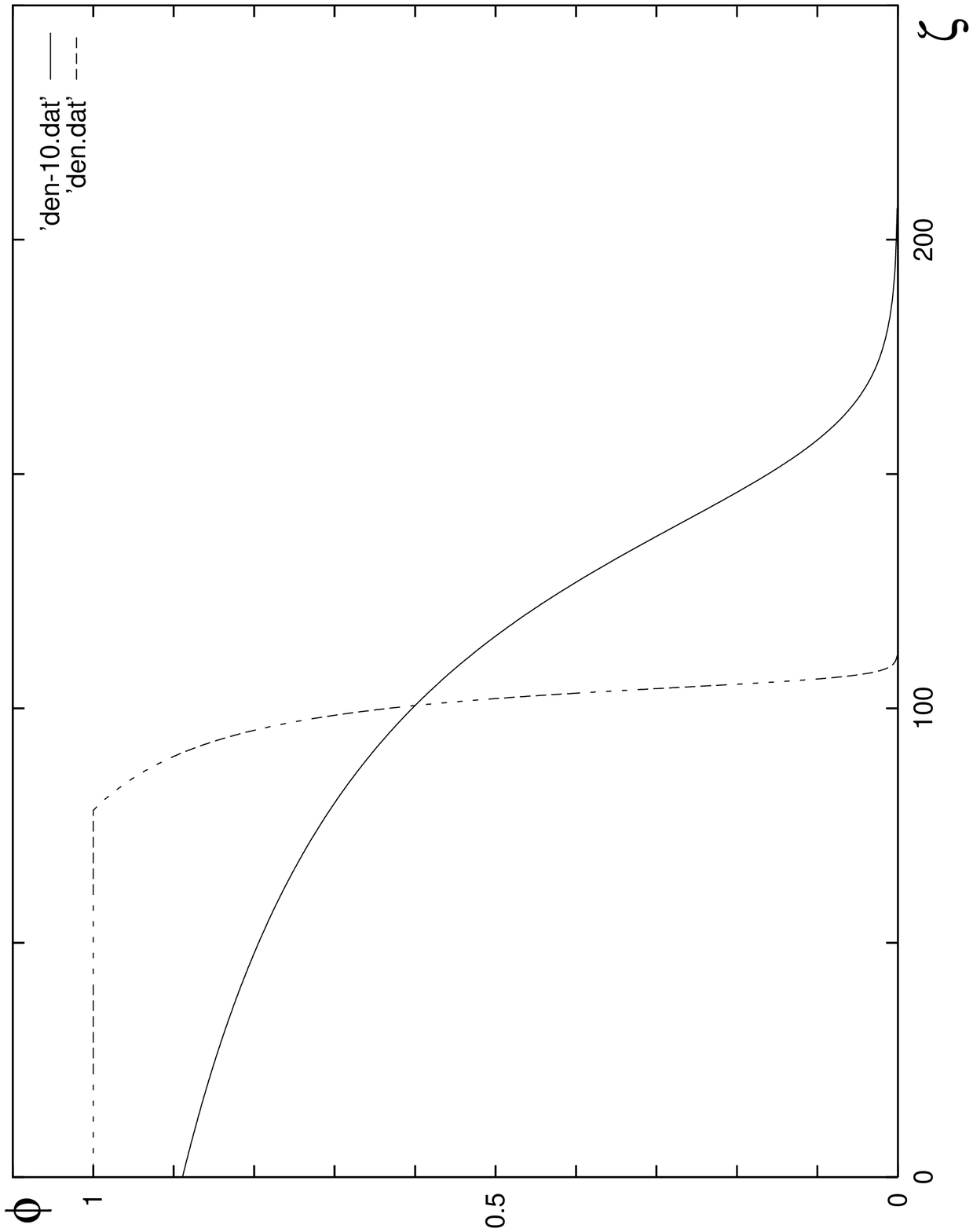}
%,height=3.5in,width=5.5in,angle=270}
}}

\end{document}